\input epsf  
\documentstyle[11pt,aaspp4]{article} 
\def\beq{\begin{equation}}
\def\eeq{\end{equation}}
\def\bey{\begin{eqnarray}}
\def\eey{\end{eqnarray}}
\def\kms{\mbox{\rm \,km\,s}^{-1}}
\input epsf
\begin{document}
\title{A Radial Velocity Survey for LMC Microlensed Sources}
\author{HongSheng Zhao \thanks{Email: hsz@strw.leidenuniv.nl}
\\Sterrewacht Leiden, Niels Bohrweg 2, 2333CA Leiden, NL}
\date{Accepted $\ldots$  
      Received $\ldots$; 
      in original form $\ldots$} 
\label{firstpage} 
\begin{abstract}
We propose a radial velocity survey with the aim to resolve the
current dispute on the LMC lensing:  in the pro-macho hypothesis 
the lenses are halo white dwarfs or machos in
general; in the pro-star hypothesis both the lenses and the sources
are stars in various observed or hypothesized structures of
the Magellanic Clouds and the Galaxy.  
Star-star lensing should prefer
sources at the backside or behind the LMC disc because lensing is most
efficient if the source is located a few kpc behind a dense screen of
stars, here the thin disc of the LMC.  This signature of self-lensing
can be looked for by a radial velocity survey since kinematics of the
stars at the back can be markedly different from that of the majority
of stars in the cold, rapidly rotating disc of the LMC.  Detailed
simulations of effect together with optimal strategies of carrying out
the proposed survey are reported here.  Assuming that
the existing 30 or so alerted stars in the LMC are
truely microlensed stars, their kinematics 
can test the two lensing scenarios; the confidence level
varies with the still very uncertain structure of the LMC.
Spectroscopy of the existing sample 
and future events requires about two or three good-seeing nights per year at 
a 4m-8m class southern telescope, either during the amplification phase
or long after.
\end{abstract}


\section{Introduction}

\begin{figure} 
\epsfxsize=10cm 
\centerline{\epsfbox{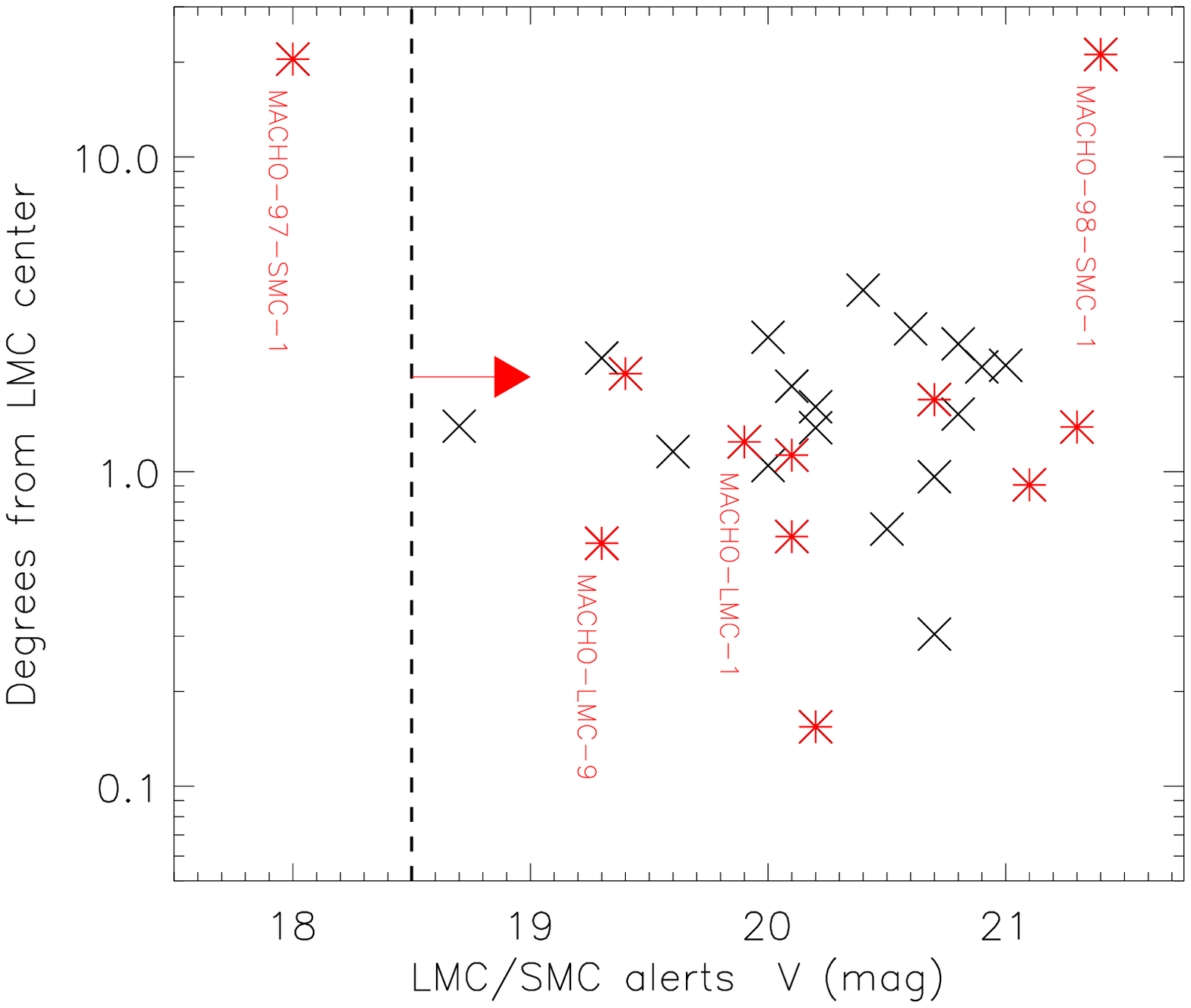}}
\caption{
Magnitude and radius distribution of some 
30 microlensing alerts (crosses) and published candidates 
(asterisks) towards the Magellanic Clouds.  The dashed line
marks the distance modulus of the LMC.
Lensed stars should be fainter in average 
(direction of arrow) compared to the unlensed stars and cluster 
at where the LMC has the biggest line of sight depth
if LMC-LMC self-lensing is important.
}\label{vmag}
\end{figure}

Experimental search for microlenses in the line of sight to the LMC by
MACHO, OGLE, EROS and a number of follow-up surveys has found more
than $16$ candidates and two towards the SMC.  Their magnitude and
angular distributions are shown in Fig.~\ref{vmag}, including some
unconfirmed alerts.  One complication to the otherwise plausible
conversion of the event rate to $\Omega_{macho}$ of the universe is
the inevitable background events, on top of any macho signal, coming
from self-lensing of stars in the LMC disc (Sahu 1994).  Since
disc-disc lensing is not very efficient if the LMC disc is cold and
thin (Gould 1994, Wu 1994), more speculative non-standard
substructures to the Magellanic Clouds (MCs) have been invoked 
in the pro-star models to boost up the star-star lensing; in particular, a
connection is drawn between the unexpected rate of microlensing and
the Milky Way-MCs and SMC-LMC interactions (Zhao 1998a, b, Weinberg
1998).

Numerous observational and theoretical arguments and counter-arguments
have been presented on the stellar distribution in the line of sight
to the MCs.  These are summarized in Zhao (1999a) and divided to two
broad classes.  In the Type-I models either the sources or the lenses
are objects of origin independent of the MCs.  For example, the lenses
could be the machos in the Galactic halo or stars in the Galactic
thick disc or warp (Evans et al. 1998).  In this case the lens and
the source most likely move with widely different velocities.  In the
Type-II models both the sources and the lenses are stars co-moving
with (within $100\kms$ of the systematic velocity of) the MCs. These
are generally from substructures which are one way or another
generated by dynamical processes in the formation and evolution of the
MCs or their progenitor.  It is debatable whether the geometry of
these substructures is better described as a thickened disc, or a
warp, a flare, a polar ring, a tidal halo etc. (Zhao 1998a,b, Weinberg
1998), but the common feature is that they circulate around the Galaxy
together with the Clouds.

The Magellanic Clouds are known to sport several peculiar
substructures for a long time.  Recent discovery of the leading arm of
the Magellanic Stream lends further weight to the old idea (Lu et
al. 1998, Putman et al. 1998 and references therein) that a recent
($\sim 0.5$ Gyr) close encounter of the SMC and the LMC is responsible
for the formation of the Magellanic Stream and the gaseous envelope
(Magellanic Bridge) connecting the MCs.  It seems promising to use
these tidal interaction models to explain, among other substructures,
the recently found polar-ring-like structure surrounding the LMC disc
traced by a radial velocity distribution of bright carbon stars
(Kunkel et al. 1997) and a faint round halo of the LMC in the USNO2
and 2MASS data (Weinberg 1998).  Nevertheless only qualitative
predictions should be trusted for the amount of stellar material
stirred up by the tidal interactions and their distribution because of
a large number of free parameters.  It is thus most interesting to
seek generic signatures of the tidal models on lensing.

Here we show that macho-LMC lensing models (or generally Type-I
models) and LMC-LMC self-lensing models (Type-II models) leave
slightly different footprints on the kinematic and spatial
distribution of the lensed sources.  A number of ways to resolve the
substructures in the line of sight have already been discussed in Zhao
(1999a).  These make use of the radial velocity, distance, proper
motion and spatial distribution of the microlensed sources.  Among
these, it seems most promising to look for a self-lensing-induced bias
of radial velocities of the sources.  Here a quantitative estimation
of this bias is given.  Our calculations of a specific set of models
should serve merely as examples of general signatures that we expect
of a wide class of models.

\section{Model of the LMC disc}

We assume a nearly flat rotation curve for the disc stars of the LMC
with of a rotation speed
\beq
V_c(r)={V_0 r\over \sqrt{r^2+r_c^2}},~~~\mbox{\rm for~~~}r\le r_t,
\eeq
where 
$r_c$ is the size of the core, and the $r_t$ is the tidal truncation radius.
On top of the rotation
an isotropic Gaussian of dispersion $\sigma$ is used to model the
random motion of the disc stars.  So the stellar distribution function
\beq
DF_d = {\rho_d(R,Z) \over (2 \pi)^{3\over 2}\sigma^3} \exp\left[-{(V_\phi-V_c)^2+V_R^2+V_Z^2 \over 2 \sigma^2}\right],
\eeq
in the cylindrical coordinates $(R,Z,\phi)$ and 
the corresponding velocity coordinates $(V_R,V_Z,V_\phi)$, where
the $Z-$axis points towards the rotation axis of the LMC disc.  
We model the LMC as an exponential thin disc of a mass $M_d$ and
a scale-length $R_d$ with a volume density
\beq
\rho_d(R,Z) = {M_d \over (2\pi)^{3/2} R_d^2Z_d} \exp\left(-{R \over R_d}\right) \exp\left(-{Z^2\over 2 Z_d^2}\right).
\eeq
We adopt $R_d=2r_c=1.6$kpc and $V_0=5\sigma=70\kms$ (Freeman et
al. 1983, Bothun \& Thompson 1988); the small dispersion of about
$15\kms$ is appropriate for the young disc of the LMC (Kunkel et
al. 1997).  The LMC disc should be truncated at about $r_t=6-10$kpc, 
outside which the 
tidal force from the Galaxy at pericenters of the LMC orbit ($\sim
40$kpc) becomes important.  Kunkel et al. found that
the total dynamical mass within the
truncation radius is about $6\times 10^9M_\odot$, much lower than 
the value commonly used in the literature, $2\times 10^{10}M_\odot$.
So there appears
to be little need to introduce a dark halo in the LMC, since
the lower mass is quite comparable to that of the disc
with a total luminosity of about $4\times 10^9L_\odot$ in the B-band
and a mass-to-light ratio of $1-2M_\odot/L_\odot$.  
To be conservative on the rate of self-lensing, 
we adopt a small and light LMC with a disc mass 
$M_d=6\times 10^9M_\odot$, truncated at $r_t=6$kpc.  

The vertical profile
of our LMC disc model is assumed to be a Gaussian with a scale-height
$Z_d$. For a disc in hydrostatic equilibrium in the vertical direction
$Z_d$ should increase with the radius $R$ if the vertical dispersion
$\sigma$ is kept constant.  Alternatively $\sigma$ should decrease
with radius for a model of a constant scale-height; we adopt the
former case since the Carbon stars of the LMC appear to be equally
cold at all radii (Kunkel et al.).  Treating the disc stars as
test particles in a spherical potential with the above rotation curve,
we find
\beq 
Z_d(R)={\sigma \over V_c(R)} R.
\eeq 
Thus our disc flares up from a scale-height $Z_d$ of
$160$pc at the center to $1.2$kpc at the tidal truncation of the disc
with a maximum opening angle of the flare of $12^o$.
The strong flaring at large radius could become important for lensing
if the LMC disc is close to edge-on, but 
cannot fully account for events observed within $3^o$ of
the LMC center for a nearly face-on geometry.

The whole disc is inclined with an angle $i_d$ from the sky plane.  A
convenient and fairly good approximation of the observed orientation
of the LMC disc is such that the disc is orthogonal to the Galactic
disc and in projection the short axis of the disc runs along the line
of constant Galactic longitude.  Since the center of the LMC is at
$(l,b)=(280^o, -33^o$), this makes the inclination $i_d=33^o$ and the
PA axis orthogonal to $l=280^o$.  The far side of the disc is at more
negative latitude, and the rotation axis points roughly away from us
with the more negative longitude side being red-shifted with a velocity
$V_0\sin i_d=38\kms$.

\section{Models of stars co-moving with the Magellanic Clouds}

The co-moving material is modeled as a uniform loop or torus of total
mass $M_t$, wrapping around the LMC disc.  We assume a circular
cross-section for the torus with an area $\pi \left({r_{out}-r_{in}
\over 2}\right)^2$ where $r_{out}$ and $r_{in}$ are the radii of the
outer and inner edges of the torus from the center.  Stars circulate
around the torus with a mean angular momentum $L_t$.  On top of the
rotation the velocity distribution is an isotropic Gaussian with 
dispersion $u$.
So the stellar distribution function $DF_t$
in cylindrical coordinates $(R',Z',\phi')$ and velocities
$(V_{R'},V_{Z'},V_{\phi'})$ is given by
\beq
DF_t= {\rho_t(R',Z') \over (2 \pi)^{3\over 2}u^3} \exp\left[-{
\left(V_{\phi'}-L_t/R'\right)^2+V_{R'}^2+V_{Z'}^2 \over 2 u^2}\right],
\eeq
where the 
$Z'-$axis is tilted from the $Z-$axis of the LMC by an angle $\beta$.
The torus has a uniform volume density
\beq
\rho_t = {4 M_t \over  \pi^2 (r_{out}^2-r_{in}^2)(r_{out}-r_{in})},
~~~\mbox{\rm for~~~}\left(R'-{r_{out}+r_{in} \over 2}\right)^2+Z'^2\le \left({r_{out}-r_{in} \over 2}\right)^2.
\eeq

Our torus might be considered as a simplified model of unvirialized
substructures in the LMC due to tidal interactions with infalling
objects over a Hubble time.  For example, the stellar halo or disc of
the LMC might form from accreted lumps and the LMC disc might be
warped or thickened.  While speculative, it is conceivable that the
LMC could well be a scaled-down version of the Milky Way, which is
known to host a number of substructures, including the Sagittarius
stream and a dozen satellite galaxies, a thick stellar disc and a
warped HI disc.  In views of these, the torus is an approximation to a
wrapped-around tidal stream in the halo of the LMC; a stream can be
pulled from a lump of stars and gas when it makes a pericentric
approach to the LMC disc; the SMC is thought to have experienced such
a destructive plunge.  It seems reasonable to set the parameters
$M_t=2\times 10^9M_\odot$, $r_{out}=6 r_{in}=6$kpc,
$L_t/r_{out}=20\kms$ and $u=30\kms$, which makes a thick torus
spanning the radii from the core of the LMC to the end of the LMC
disc.  The total mass and velocity FWHM of the torus are 
equal to those of the SMC, which could be argued as the most massive
satellite of the LMC.  We vary the
angle of the tilt $\beta$ for the torus so that its geometry varies
from that of a thickened disc co-planar with the LMC disc ($i_t=i_d$
and $\beta=0^o$), to that of a polar ring of the LMC edge-on to us
($i_t=\beta=90^o$).

\section{Event rate and distributions}

Finally we need to assume a mass and luminosity function for the LMC
disc and the co-moving material before simulating the microlensing
event rate of these models.  We assume that
the phase space number density, 
in unit of $({\rm pc} \times \kms)^{-3}$, of objects with mass $m$ is given by
\beq
{d n \over d \log m} = {DF \over M_\odot}\times {\rm Max}\left(1, {m \over 0.5M_\odot}\right)^{-1.75},~~~\mbox{\rm for~~~} 0 \le m\le 3M_\odot,
\eeq
where ${DF \over M_\odot}$ is the phase space number density of the lenses 
integrated over the lens mass $m$.
The shape of the mass spectrum is a power-law close to Salpeter slope 
for massive stars and flattens out for stars and brown dwarfs 
below half a solar mass; this
is roughly consistent with observed CMDs with Hubble Space
Telescope of the LMC, solar neighbourhood and Galactic bulge 
(Holtzman et al. 1997, 1998, Gould et al. 1997, Kroupa, Tout \& Gilmore 1993).

The microlensing event rate, $\Gamma$ per observable star
per year, can then be calculated with
\beq
\Gamma = {N_l \over N_* T},~~~\mbox{\rm where~~~}
N_*= \int \eta_s n_s \Omega D_s^2 dD_s dV_s^3
\eeq
is the number of observable stars in a solid angle $\Omega$ of a survey, and
\beq\label{nl}
N_l = \int d N_* \eta \left({dn_l \over d\log m} d\log m \right) 
\left(|\mu_s-\mu_l|T \times 2 \theta_E \right) 
D_l^2 d D_l dV_l^3,
\eeq
is the number of detected lenses for a survey of duration $T$,
\beq\label{einstein}
2\theta_E=|\mu_s-\mu_l|\hat{t} = 4 \sqrt{G\,m (D_s-D_l) \over c^2 D_s D_l} 
\eeq
is the Einstein ring angular diameter for a lens with mass $m$,
$\hat{t}$ is the time to cross this diameter if $D_l$, $\mu_l$, and
$V_l$ are the distance, the proper motion vector, the 3D velocity
vector of the lens, and $D_s$, $\mu_s$, and $V_s$those of the source.
The phase space densities of lens and source are $n_l$ and $n_s$.  We
assign $\eta_l$ and $\eta_s$ as the detection efficiency for an event
of duration $\hat{t}$ and the selection function of the source defined
by the survey magnitude limit.  Note that the lensing
of a given source is more effective for increasing
lens density $n_l$ and source-lens distance 
$D_s-D_l$ (cf. eq.~\ref{nl} and~\ref{einstein}). We shall simulate the events
for a theoretical microlensing survey of $\Omega=1' \times 1'$
field for $T=10^6$ years, the equivalent of $100$
square degrees (the entire LMC) for $3$ years in terms of total exposure.  
Our calculations are
done assuming the detection efficiency of the MACHO survey and that
every one star per $40M_\odot$ of the LMC is an observable target,
approximately the case for stars above $20$mag.

\section{Results}

\begin{figure} 
\epsfysize=22cm 
\epsfxsize=18cm 
\centerline{\epsfbox{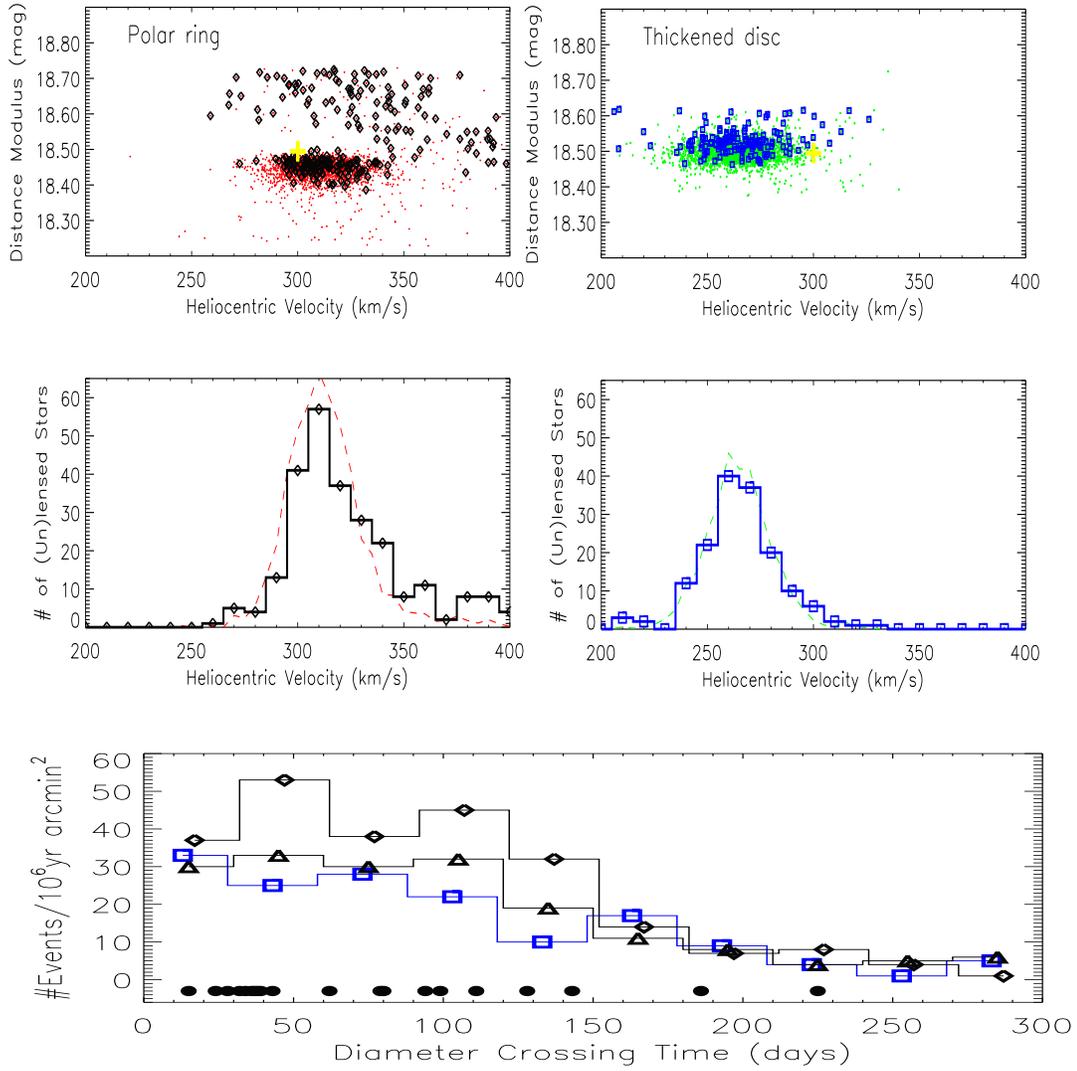}} \vskip -4cm
\caption{
Upper panels: radial velocity vs. distance modulus distribution
for the lensed sources in a given line of sight (diamonds for a minor axis
field of the polar ring model, and squares a major axis field of the
thickened disc model).  Note they
bias towards kinematically distinct stars behind the LMC
disc, unlike unlensed random stars (small dots) in the same direction.
The large crosses mark the systematic velocity and distance of the LMC.
Middle panels: histograms of the radial velocity
of lensed sources (solid lines with symbols) and 
a renormalized distribution for neighbouring unlensed stars (dashed lines).
Note the wings of lensed sources.
Bottom panels: histograms of predicted event time scale in bins of 1 month;
the one with triangles is for a major axis field of the polar ring model.
A few MACHO alerts are indicated by solid circles at the bottom.
}
\label{vrdm}
\end{figure}

The upper panels of Fig.~\ref{vrdm} show the simulated velocity and
distance distributions for the polar ring model in a line of sight on
the minor axis of the LMC $(283^o, -33^o)$, and for the thickened disc
model on the major axis $(280^o, -31^o)$.  They are typical lines of
sight of MACHO observed events and are chosen to have nearly the same
number ($\sim 2000$) of LMC disc stars per square arcmin.

There are mainly two complementing effects shown here.  First a radial
velocity survey should pick up a small fraction of outliers of the
rotation curve of the LMC disc, which may belong to some puffed-up
distribution (be it a polar ring or a thickened disc) surrounding the
LMC disc.  Second the outliers at the backside of the LMC are more
likely picked as lensed sources.  {\it They show up as the wings of
the velocity histogram with a velocity set apart from the rotation
speed of the LMC disc in the same field by typically more than
$20{\mbox{\rm \,km\,s}}^{-1}$.}  The velocity distribution for the
lensed sources has the strongest wings for the minor axis field of the
polar ring model.  The distribution has a probability of $10^{-17}$
(from an F-test of the variance) being the same as the distribution of
the unlensed stars in the LMC.  In contrast, if all events come from
LMC disc stars being lensed by foreground Milky Way machos or disc
stars, then the lensed sources would follow the motions of average
stars in the cold, rotating disc.  The confidence level here should not
be taken literally given systematic uncertainties with
the underlying models of the LMC.

We have run several models, varying the angle between the torus and
the LMC disc.  Occasionally the wings disappear even in self-lensing
models, in which case the degeneracy with the macho-lensing models
remains.  This, for example, happens when the line of sight misses the
torus-like substructure.  But in general the high velocity wings
appear to be generic and significant (cf. Fig.~\ref{vrdm}).  The
F-test gives a probability of $6\times 10^{-4}$ or $4\times 10^{-12}$
respectively for the major axis field ($+3^o$ from center) of the
polar ring model or the thickened disc model.  The above statistics
are all computed for samples of a few hundred events, which is
relevant for a future microlensing survey of the LMC.  But even for
the present, much smaller sample of $10-30$ lensing events the wings
in the velocity distribution should still be at a level marginally (by
about $1\sigma$) observable if self-lensing is indeed dominant.

There is one interesting variation of the polar ring model.  Generally
the ring is not exactly edge-on, as it has been assumed, so some LMC
stars will be lensing polar ring stars at the back in one strip of the
sky, and being lensed by polar ring stars in the front in another
strip of the sky.  Events will then be split equally between the two
configurations, as it is with the thickened disc model.  However, the
phase-mixing of a tidal substructure will be incomplete even over a
Hubble time, so a polar stream might appear only at the backside of
the LMC disc.  Such a polar stream or ``arc'' should show up as a
strong signal in the velocity distribution of the lensed sources and
in the sky distribution of the events.

Finally we comment briefly on the event rate and event time scale
distribution (the bottom panel of Fig.~\ref{vrdm}).  
First note the interesting signature of self-lensing:
the event duration distribution and event rate vary across the LMC
because the typical lens-source relative distance and velocity all vary with
line of sight.
Also the long duration tails of the event distributions are generally
strong because the lenses are stars with mass 
above the hydrogen burning limit, and the lens and the source 
are co-moving with relative
velocity limited by the escape velocity from the LMC potential.
Second, there appears to be a general shortage of stars in
co-moving substructure of the LMC available as lenses
or sources.   The lensing rate
for the polar ring model is about $0.08 \times (M_t/2\times
10^9M_\odot)f_{\rm wrap}^{-1}$ and 
$0.11 \times (M_t/2\times 10^9M_\odot)f_{\rm wrap}^{-1}$ per survey
star per million year on the major axis and minor axis respectively,
where $M_t$ is the mass of the 3D substructure,
and the factor $f_{\rm wrap}$ takes into account that an unviralized
substructure may not wrap around a full circle of the LMC, but only
a fraction $f_{\rm wrap}$ of it.
These event rates appear significantly
(by a factor three) lower than the observed rate
if we believe that the LMC disc is thin ($\le 400$pc in
FWHM), the 3-dimensional substructure (whether a polar
stream or a thickened disc component) is a light ($M_t \le 2\times
10^9M_\odot$) and wrapped-around symmetric component ($f_{\rm wrap}=1$), and
the tidal truncation radius of both is small ($r_t \sim 6$kpc).
  To boost-up the self-lensing rate to account for
all observed events, it appears necessary to (a) invoke a stubby 
($f_{\rm wrap}\sim 0.1-0.5$), 
SMC-sized, tidal stream of the LMC lying at a few kpc either in the
back or front of the LMC disc (Zhao 1998a,b), or (b) put 
most of the LMC's mass in a thickened component
($M_t \sim 10^{10}M_\odot$) out to $10$kpc from the center of the LMC
(Weinberg 1998).  
The proposed kinematic survey should place stringent limit on
the fraction of older and kinematically distinct 
populations in the LMC disc and
surrounding.

\section{Strategy for the radial velocity survey}

Most of current alerts and candidates of lensed source stars are about $(20\pm
1)$mag long after the microlensing event (cf. Fig.~\ref{vmag}), and
brighter than 19 mag at the peak of the amplification.  They are
bright enough for obtaining low resolution spectrum using 4m-8m class
southern telescopes.  The observations can be scheduled any time after
the event.  To keep up with the current turn-out rate of microlenses
towards the LMC and SMC ($\le 10$ events per year), we need about two
nights each year during the latter half of the LMC/SMC season with a
single/multi-slit spectrograph.  As an example, we estimate that a
ten-minute integration with the VLT can achieve a S/N of 10 per \AA,
sufficient for the radial velocity work here, for a $V=20$mag star
using FORS1 in its highest resolution mode.  
Comparable S/N might be achieved with an one-hour exposure on
other 4m class telescopes in Chile and Australia (e.g., Sahu \& Sahu
1998) depending on seeing and instruments.  
While taking spectrum at the peak can obviously achieve the
same S/N for less integration time (Lenon et al. 1996), it is probably
not advisable considering its heavy demand for immediate observing.

A continuation and expansion of the MACHO survey, such as the Next
Generation Microlensing Survey (NGMS, Stubbs 1998), holds the promise
to raise the event turn-out rate by an order of magnitude.  This
involves monitoring the light curves of ten times more stars than the
9 million by the MACHO program.  It effectively includes if all stars
brighter than 20-21 mag. over the entire LMC.  To follow up these
events, a more ambitious program appears to be necessary.  It is
probably optimal for a 2m class telescope (such as the Du Pont 2.5m at 
Las Campanas) with a single-slit or
multi-object capability to take spectra of every event near the peak
since we expect a new source coming to maximum amplification every few
nights; dedicated use during the LMC/SMC season would be most
efficient.  The reduced aperture is nicely compensated by the
gravitational amplification; a $1.5$ magnitude amplification is
roughly equivalent to doubling the aperture of the telescope.  The
gain of taking spectra in real time would be even greater for the rare
class of high amplification events as done for MACHO-96-BULGE-3 and
MACHO-98-SMC-1.  The size of the source star can be inferred from the
spectra, which could constrain the relative proper motion of the lens
and source $|\mu_l-\mu_s|$.  Very strong constraints on the lensing
model can come from a combination of radial velocity and proper motion
(cf. Fig.~1 of Zhao 1999a).

Another integral part of the survey is to obtain radial velocities of
random stars in the LMC, particular those in the immediate
neighbourhood of the lensed stars.  These neighbouring stars serve
both as a direct probe of the hotness of the LMC and existences of any
3D substructure and as a control sample quantifying the velocity wings
of the microlensed sources.  All together, the kinematic survey can be
made most efficient by the following.  For every microlensed source
under study we can, at the same time, take spectra of $10$ (say with
VLT/FORS1) to $100$ (say with AAT/2dF) well-isolated stars in the same
observing field depending on the crowdness of the field and the
capability of the multi-object spectrograph.  The candidate list for
spectra can be built from CMDs for identifying the lensed source in
the first place, and only those field stars brighter than the target
source star should be used for efficiency.  With this strategy we
expect to build a sample of $10^3-10^4$ stars in the vicinity of
10-100 lensed stars.  The large sample is necessary to be sensitive to
minor components; the VRC stars of Zaritsky et al. (1997, 1999) make
up only a few percent of the total luminosity of the LMC.  A fair
sample should also include stars of a wide range of intrinsic
luminosity and color so as to minimizing biasing due to different
ages; the older population might be more puffed-up and
rotate slower than the young, thin disc.  As a by-product of the dark
matter study, these kinematic samples are useful for understanding of
the spatial and dynamical structure of the LMC.

\section{Conclusion}

Subtle systematic effects are expected for microlensed stars towards
the LMC in their distributions of radial velocity, proper motion, 
projected distance from the center of the LMC, 
distance modulus and reddening
(Zhao 1999a,b).  While the predictions are still subject to model
assumptions at different levels, detection or non-detection of these
lensing-induced systematic bias in all five distributions is likely to
give a robust conclusion on the nature of Galactic dark matter.  The
radial velocity survey is by far the most promising approach since
radial velocities can be measured very accurately with access to 2m-8m
southern telescopes.  The survey can be done for both exotic events
and common low amplification events either during lensing or at a
scheduled time in the latter part of the season.  Studies of the
systematic bias can all be integrated in the Next Generation
Microlensing Survey (Stubbs 1998) for the dark matter problem and
other photometric/kinematic surveys of the LMC for the studies of
structures and star formation history of the LMC (Zaritsky et
al. 1999).  

Table 1 shows the levels of confidence to rule out macho-lensing
models (or other Type-I models with intervening material dynamically
unassociated with the LMC) with the current sample and a larger sample
of microlensing events in the future, assuming that we will obtain the
radial velocities of all the lensed sources and unlensed stars in
neighbouring lines of sight (say, within $5'$, which is 100pc at the
LMC).  More than half a decade of the MACHO survey have covered more
than $11$ square degrees and $9$ million stars of the LMC and SMC, and
found up to 30 possible microlensing events.  The total exposure is
about $(0.5-1)\times 10^8$ in units of star year.  We expect that the
exposure and the number of events will be more than quadrapoled in the
next few years by the on-going EROS 2 and OGLE 2 surveys, the MACHO
last-year survey and some versions of NGMS.  While spectra of
microlensed sources have been taken sporadicly in the past (e.g. Della
Valle 1994 on MACHO-LMC-1), there is a lack of differential study of
kinematics and photometry with unlensed stars in the same field.  As
of present there is no published data on the kinematics of the
microlensed sources.

Finally these differential studies of microlensing stars and unlensed
stars could also be done towards the Galactic bulge where self-lensing
of the Galactic bar surely plays an important role.  The Space
Interferometry Mission (SIM, 2005-2010) provides a more distant but
exciting prospect of measuring the lensing-induced wobbling motion of
the image centroid as astrometric follow-up of a few microlensing
candidates from ground-based surveys.

The author thanks Wyn Evans, Ken Freeman, Puragra Guhathakurta,
Rodrigo Ibata, Konrad Kuijken, Knut Olsen, Joel Primack, Penny Sackett
and Chris Stubbs for helpful discussions and the anonymous referee for
very helpful comments on the presentation.

\begin{deluxetable}{lrrrr}
\tablewidth{0pc}
\tablecaption{Sample size and confidence level for discriminating 
various microlensing hypotheses}
\tablehead{
  \colhead{Exposure}\tablenotemark{1}
& \colhead{$N_{\rm source}$}\tablenotemark{2} 
& \colhead{$N_{\rm neighbour}$}\tablenotemark{3} 
& \colhead{Confidence}\tablenotemark{4} 
& \colhead{Confidence}\tablenotemark{5}
}
\startdata
 0.5e+08 &    25 &   500 & 2.e-02 &      4.e-02 \nl
 1.5e+08 &    75 &  1500 & 3.e-05 &      6.e-04 \nl
 2.5e+08 &   125 &  2500 & 9.e-08 &      9.e-06 \nl
 3.5e+08 &   175 &  3500 & 3.e-10 &      2.e-07 \nl
 4.5e+08 &   225 &  4500 & 8.e-13 &      3.e-09 \nl
 5.5e+08 &   275 &  5500 & 3.e-15 &      5.e-11 \nl
\enddata
\tablenotetext{1}{The exposure, 
which is a product of the number of stars monitored
and the duration of the microlensing survey.}
\tablenotetext{2}{The number of lensed sources expected.}
\tablenotetext{3}{The number of unlensed bright stars 
in neighbouring (within $5'$) lines of sight,
available as targets for the proposed differential study of radial velocities.}
\tablenotetext{4}{The F-test confidence levels of ruling out macho-lensing 
with this kinematic study if in reality the events are 
all drawn from our LMC thin disc plus a polar-ring model.  
These confidence levels are merely illustrative since
the models of the LMC are far from unique.}
\tablenotetext{5}{Same as (4) except that events are all drawn from 
our LMC thin disc plus a thickened disc model.}
\end{deluxetable}

{}

\vfill \eject

\end{document}